\documentclass[structabstract]{aa}

\usepackage[utf8]{inputenc}
\usepackage{graphicx}
\usepackage{natbib}
\usepackage{xcolor}
\usepackage{hyperref}
\usepackage{txfonts}
\usepackage{longtable}

\title{Observed tidal evolution of Kleopatra's outer satellite
\thanks{Based on observations made with ESO Telescopes
at the La Silla Paranal Observatory under program 199.C-0074 (PI Vernazza).}
}

\titlerunning{Observed tidal evolution of Kleopatra's outer satellite}

\author{
    M.~Bro\v{z}\inst{\ref{prague}}         \and 
    J.~\v{D}urech\inst{\ref{prague}}      \and 
    B.~Carry\inst{\ref{oca}}               \and 
    F.~Vachier\inst{\ref{imcce}}           \and 
    F.~Marchis\inst{\ref{seti},\ref{lam}}  \and 
    J.~Hanu{\v s}\inst{\ref{prague}}       \and 
    L.~Jorda\inst{\ref{lam}}               \and 
    P.~Vernazza\inst{\ref{lam}}            \and 
    D.~Vokrouhlick\'y\inst{\ref{prague}}   \and 
    M.~Walterov\'a\inst{\ref{dlr}}         \and 
    R.~Behrend\inst{\ref{behrend}}          
}

   \institute{
     Institute of Astronomy, Faculty of Mathematics and Physics, Charles University, V~Hole{\v s}ovi{\v c}k{\'a}ch 2, 18000 Prague, Czech Republic%
     \label{prague}%
     \and 
     Universit\'e C{\^o}te d'Azur, Observatoire de la C{\^o}te d'Azur, CNRS, Laboratoire Lagrange, France
     \label{oca}
     \and 
     IMCCE, Observatoire de Paris, PSL Research University, CNRS, Sorbonne Universit{\'e}s, UPMC Univ. Paris 06, Univ. Lille, France%
     \label{imcce}
     \and 
     SETI Institute, Carl Sagan Center, 189 Bernado Avenue, Mountain View CA 94043, USA 
     \label{seti}
     \and 
     Aix Marseille Univ, CNRS, LAM, Laboratoire d'Astrophysique de Marseille, Marseille, France
     \label{lam}
     \and 
     Institute of Planetary Research, German Aerospace Center (DLR), Rutherfordstr. 2, 12489 Berlin, Germany
     \label{dlr}
     \and 
     Geneva Observatory, CH-1290 Sauverny, Switzerland
     \label{behrend}
}

\date{Received x-x-2021 / Accepted x-x-2021}

\abstract
   {}
   {
The orbit of the outer satellite Alexhelios of (216) Kleopatra
is already constrained by adaptive-optics astrometry,
obtained with the VLT/SPHERE instrument.
However, there is also a preceding occultation event in 1980
attributed to this satellite.
Hereinafter, we try to link all observations, spanning 1980--2018.
We find the nominal orbit exhibits an unexplained shift by $+60^\circ$
in the true longitude.
   } 
   {
Using both periodogram analysis and an $\ell = 10$ multipole model
suitable for the motion of mutually interacting moons
about the irregular body,
we confirmed that it is not possible to adjust
the respective osculating period $P_2$.
Instead, we were forced to use a model with tidal dissipation
(and increasing orbital periods) to explain the shift.
We also analysed light curves, spanning 1977--2021,
and searched for the expected spin deceleration of Kleopatra.
   }
   {
According to our best-fit model,
the observed period rate is
$\dot P_2 = (1.8\pm 0.1)\cdot 10^{-8}\,{\rm d}\,{\rm d}^{-1}$
and the corresponding time lag
$\Delta t_2 = 42\,{\rm s}$ of tides,
for the assumed value of the Love number
$k_2 = 0.3$.
It is the first detection of tidal evolution
for moons orbiting 100-km asteroids.
The corresponding dissipation factor~$Q$
is comparable with other terrestrial bodies,
albeit at a higher loading frequency $2|\omega-n|$.
We also predict a secular evolution of the inner moon,
$\dot P_1 = 5.0\cdot 10^{-8}$,
as well as a spin deceleration of Kleopatra,
$\dot P_0 = 1.9\cdot 10^{-12}$.
In alternative models,
with moons captured in the 3:2 mean-motion resonance
or more massive moons,
the respective values of $\Delta t_2$ are a factor of 2--3 lower.
Future astrometric observations by direct imaging or occultations
should allow to distinguish between these models,
which is important for the internal structure
and mechanical properties of (216) Kleopatra.
   }
   {}
 

\begin{document}

\maketitle

\section{Introduction}

It is already known that small (1-km) binary asteroids
are driven by radiative torques, tides, or both (e.g., \citealt{Scheirich_2021Icar..36014321S}).
In case of binaries, the secondary orbital evolution is measured by means of eclipses,
as a steady decrease or increase of its period.
The primary rotation evolution is not observed though.

For large (100-km) asteroids with relatively small satellites,
the situation is different. Radiative torques (cryptographically, `BYORP')
are considered weak because they scale as
\citep{Cuk_2005Icar..176..418C}:
\begin{equation}
{\Gamma\over L} \simeq 3.0\cdot10^{-12}\,{\rm s}^{-1}\,\left({a_{\rm h}\over a_{\rm h0}}\right)^{-2} \left({\rho\over\rho_0}\right)^{-1} \left({a_1\over a_{10}}\right)^{-1} \left({R_2\over R_{20}}\right)^{-1} {P_1\over P_{10}}\,,\label{Gamma_L_radiative}
\end{equation}
where
$\Gamma$~denotes the torque,
$L$~angular momentum,
$a_{\rm h}$~heliocentric semimajor axis,
$\rho$~density,
$a_1$~binary semimajor axis,
$P_1$~its orbital period,
$R_2$~secondary radius.
The normalisation is given for
$a_{\rm h0} = 1\,{\rm au}$,
$\rho_0 = 1750\,{\rm kg}\,{\rm m}^{-3}$,
$a_{10} = 2\,{\rm km}$,
$R_{20} = 0.15\,{\rm km}$,
$P_{10} = 20\,{\rm h}$,
and synchronous rotation.

On contrary, tides scale as \citep{dePater_2010plsc.book.....D}:
\begin{equation}
{\Gamma\over L} \simeq {3\over 2}{k_2\over Q}{Gm_2^2 R_1^5\over a^6} \left({m_1m_2\over m_1+m_2}\sqrt{G(m_1+m_2) a}\right)^{-1} \,,\label{Gamma_L_tidal}
\end{equation}
where
$k_2$~denotes the Love number,
$Q$~quality factor,
$R_1$~primary radius,
$m_1$, $m_2$ component masses.
Of course, tides are known to operate in planet--moon systems,
where dissipation occurs inside the planet.
Most precisely, they are measured for the Earth--Moon system,
where
$\dot P_0 = 5.4\cdot10^{-13}\,{\rm d}\,{\rm d}^{-1}$
(primary rotation period),
$\dot P_1 = 1.1\cdot10^{-11}$
(secondary orbital period; equivalent to Moon's orbit
expansion $0.038\,{\rm m}\,{\rm y}^{-1}$).
Sometimes, dissipation must occur in the moon to explain
observed orbits (e.g., Phobos; \citealt{Rosenblatt_2011A&ARv..19...44R})
or volcanism (Io; \citealt{Peale_1979Sci...203..892P,Morabito_1979Sci...204..972M}).
There is no reason why 100-km asteroids should be different,
except for their material properties.
Unfortunately, no such measurements exist for their moons.

In this paper, we focus on the (216) Kleopatra moon system
\citep{Ostro2000,Descamps_etal_2011Icar..211.1022D,Hirabayashi_Scheeres_2014ApJ...780..160H,Shepard_etal_2018Icar..311..197S,Marchis_2021A&A...653A..57M,Broz_2021A&A...653A..56B}.
On Oct 10th 1980, an occultation of Kleopatra itself was
observed, together with a serendipitous occultation event,
which was later attributed to the outer moon of Kleopatra,
designated S/2008 (216) 1, or I~Alexhelios
\citep{Descamps_etal_2011Icar..211.1022D}.
The event took only 0.9\,s,
but was observed by two independent observers,
separated by 0.61\,km.
Its sky position in the $(u,v)$ plane coincided with the respective orbit
of the outer (2nd) moon.

\begin{figure}
\centering
\begin{tabular}{c}
\kern.5cm no tides \\
\includegraphics[width=8.5cm]{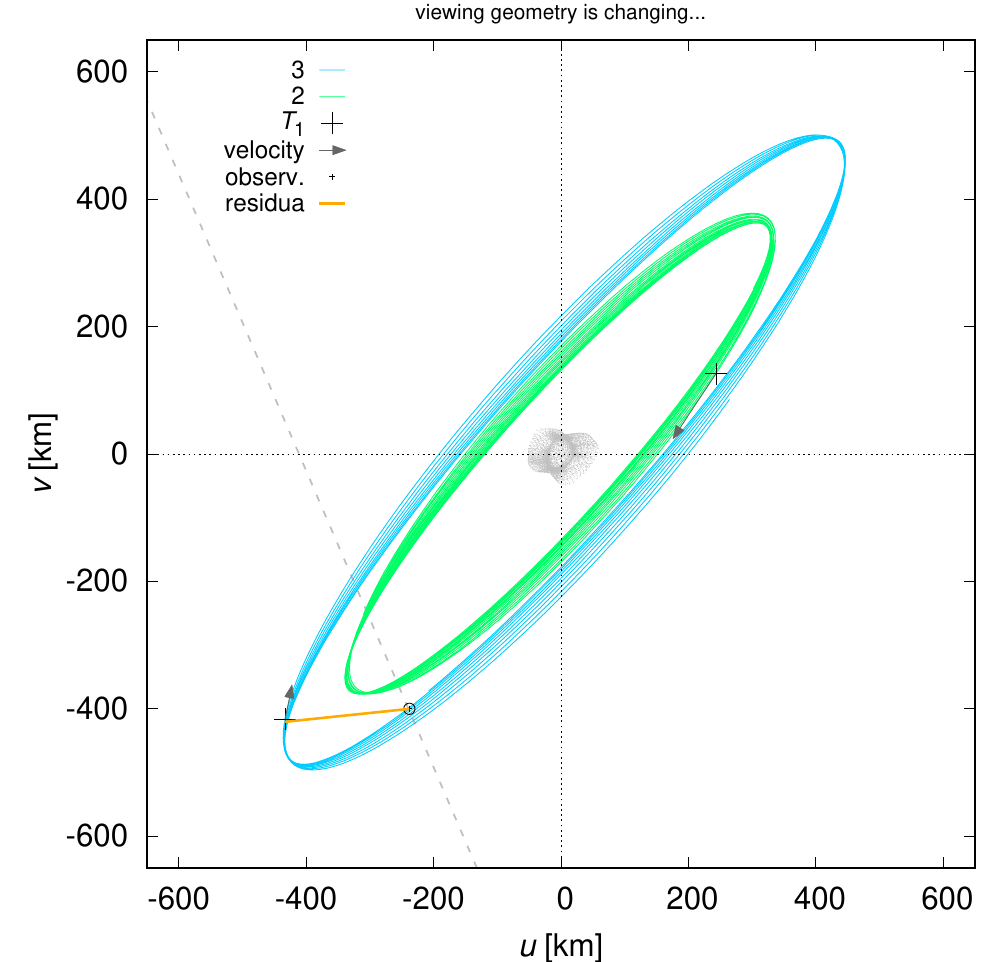} \\
\end{tabular}
\caption{Sky-plane projection of moons orbiting (216) Kleopatra,
with the observed position of Oct 10th 1980 occultation (black circle)
and corresponding chord (dashed line) from
\cite{Descamps_etal_2011Icar..211.1022D}.
For comparison both inner and outer moon orbits are plotted
(\color{green}green\color{black},
\color{blue}blue\color{black};
bodies 2, 3).
The projected orbital velocity is indicated by an arrow.
The ephemeris with constant osculating periods,
derived from adaptive-optics datasets (2008 to 2018),
is offset by ${\sim}\,60^\circ$ in the true longitude~$\lambda_2$ (black cross, orange line).
It corresponds to a shorter orbital period~$P_2$ in the past.
}
\label{216_test57_SKYANDTELESCOPE_chi2_SKY_uv}
\end{figure}

When we compared this observation with the revised ephemeris of \cite{Broz_2021A&A...653A..56B}
--- constrained by adaptive-optics (AO) datasets, hereinafter denoted as DESCAMPS, SPHERE2017, SPHERE2018 ---
it turned out that the orbit orientation is very similar,
but the predicted position is offset in the true longitude~$\lambda_2$
by approximately $+60^\circ$ (see Fig.~\ref{216_test57_SKYANDTELESCOPE_chi2_SKY_uv}).
The synthetic moon is farther away on its orbit.
This certainly requires an additional analysis,
because it can be related to tides.

The occultation can be hardly associated with the inner (1st) moon,
because the distance between the sky-plane position and the orbit
is more than $4.5\sigma$ at any given time,
and the actual longitude~$\lambda_1$ is offset
in the opposite direction by $-90^\circ$
(alternatively, by as much as $+270^\circ$).

\def\oldvec{\mathaccent"017E }
\newcommand{\xitau}{{\tt Xitau}}

\section{Observed tidal evolution}

\subsection{Increasing orbital period~$P_2$}

Na\"\i vely, we expected that a minor change of the osculating period $P_2$
within the present uncertainty
will be sufficient, but it was not. Indeed, the time span of
the AO datasets (2008--2018, or 3780\,d) is comparable with
the preceding occultation (1980--2008, 10220\,d).
Moreover, their phase coverage constrains both periods $P_1$, $P_2$.

\begin{figure}
\centering
\includegraphics[width=9cm]{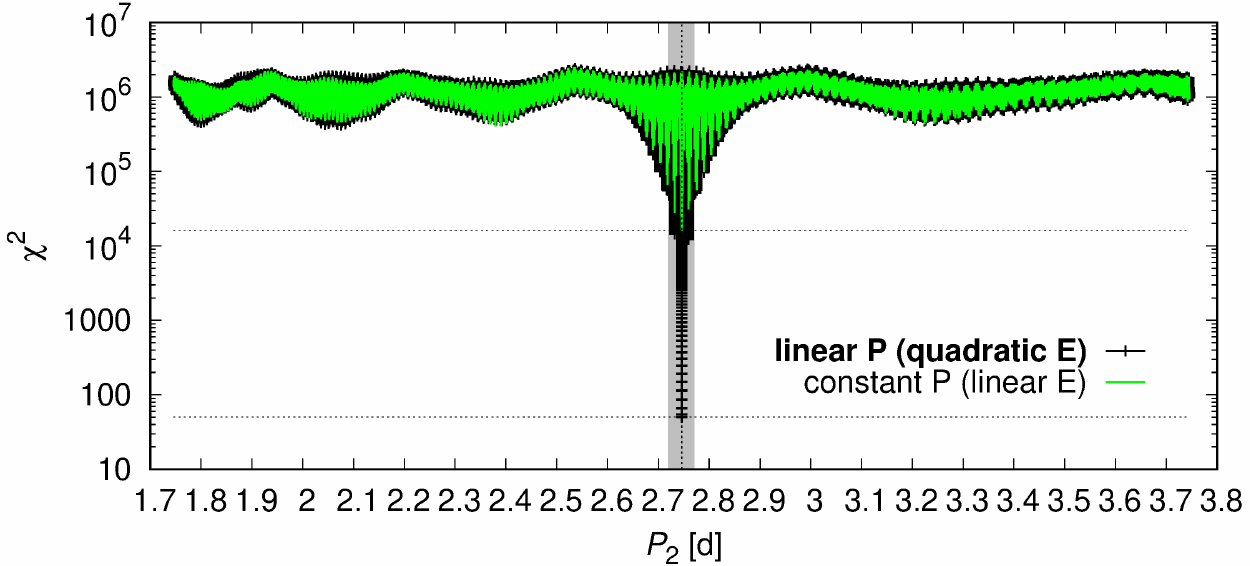}
\includegraphics[width=9cm]{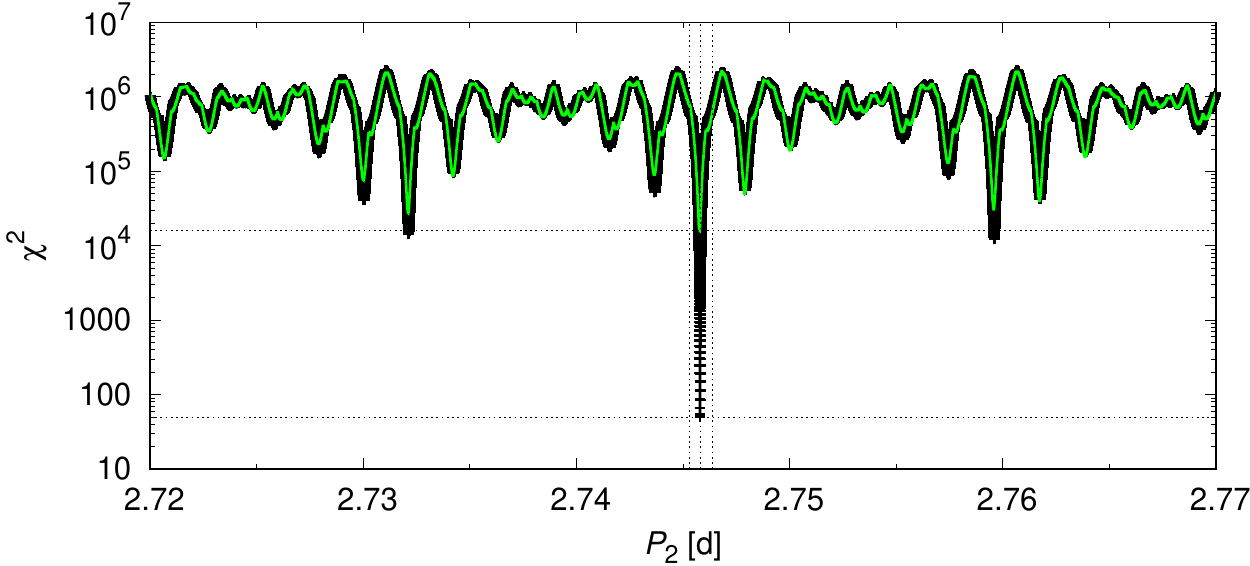}
\caption{Simplified periodograms for the 2nd moon,
obtained as a $\chi^2$ difference between the observed epochs $E_i$
and computed epochs $E(t)$
for constant mean periods~$P_2$ (green line),
and linearly variable periods $P_2(t) = P_2(0) + \dot P_2 t$ (black line).
The value $\dot P_2 = 1.8\cdot10^{-8}\,{\rm d}\,{\rm d}^{-1}$ corresponds to the offset of $\lambda_2$
in Fig.~\ref{216_test57_SKYANDTELESCOPE_chi2_SKY_uv}.
The gray box in the upper panel shows a range of the bottom panel.
}
\label{map_P2_fitphase0_1.8e-8_ALL}
\end{figure}

To demonstrate it clearly, we computed simplified periodograms
as follows. We used our previous converged model \citep{Broz_2021A&A...653A..56B}
to determine the true longitudes $\lambda_2$ (unfolded)
and orbital epochs $E_i$ of all 2008--2018 observations,
with respect to $T_0 = 2454728.761806$.
Then we added one point corresponding to the 1980 occultation,
with the respective epoch $E_i = \lambda_2/(2\pi) = 0.55$.
We assumed uncertainties $\sigma_{\kern-.5pt E} = 0.001$;
it corresponds to the astrometric uncertainty of about 10\,mas.
These data were compared with two simplified ephemerides ---
constant mean period%
\footnote{These mean keplerian periods are different from osculating periods
reported in \cite{Broz_2021A&A...653A..56B} by a factor of approximately $1.02246$.}
(linear epoch):
\begin{equation}
E(t) = {t\over P_2}\,,
\end{equation}
or linear period (quadratic epoch):
\begin{equation}
E(t) = {1\over \dot P_2}\ln\left({1 + {\dot P_2\over P_2} t}\right) \doteq {t\over P_2} - {1\over 2P_2^2}\dot P_2 t^2\,.
\end{equation}
The difference between $E_i$, $E(t)$ expressed as $\chi^2$
is plotted in Fig.~\ref{map_P2_fitphase0_1.8e-8_ALL}.
It is not possible to fit all epochs $E_i$ with any of the constant periods.
The structure of the periodograms is determined by the AO datasets,
not by the occultation.
On the other hand, a linearly variable period,
with a suitable derivative $\dot P_2 = (1.8\pm0.1)\cdot10^{-8}\,{\rm d}\,{\rm d}^{-1}$,
is satisfactory (and better by 2 orders of magnitude).

\subsection{Monopole model including tides}\label{sec:monopole}

Tidal dissipation in Kleopatra is a likely dynamical mechanism
explaining the secular evolution of the orbital period~$P_2$.
To determine basic parameters of tides, we used a time lag model
\citep{Mignard_1979M&P....20..301M,NerondeSurgy_1997A&A...318..975N}.
The additional acceleration (and torque)
was implemented in the SWIFT integrator
\citep{Levison_Duncan_1994Icar..108...18L} as follows:
\begin{equation}
\vec f_{\rm tides} = K_1 \left[K_2\vec r' - K_3\vec r - K_4 (\vec r\times\oldvec\omega + \vec v) + K_5(K_6\vec r - K_7\vec r')\right]\,,\label{eq:f_tides}
\end{equation}
\begin{equation}
K_1 = {3 Gm^\star R^5 k_2 \Delta t\over (r' r)^5}\,,\label{K1}
\end{equation}
\begin{equation}
K_2 = {5\over r'^2} \left[\vec r'\!\!\cdot\!\vec r\,(\vec r\cdot\oldvec\omega\times\vec r'\! + \vec r'\!\cdot\vec v)
- {1\over 2r^2}\vec r\cdot\vec v (5(\vec r'\!\!\cdot\!\vec r)^2 - {r'}^2 r^2)\right],
\end{equation}
\begin{equation}
K_3 = \vec r\cdot\oldvec\omega\times\vec r' + \vec r'\cdot\vec v\,,
\end{equation}
\begin{equation}
K_4 = \vec r'\cdot\vec r\,,
\end{equation}
\begin{equation}
K_5 = {\vec r\cdot\vec v\over r^2}\,,
\end{equation}
\begin{equation}
K_6 = 5\vec r'\cdot\vec r\,,
\end{equation}
\begin{equation}
K_7 = r^2\,,
\end{equation}
\begin{equation}
\oldvec\Gamma = \vec r\times m'\vec f_{\rm tides}\,.\label{eq:Gamma}
\end{equation}
The classical notation assumes Earth--Moon--test particle,
but of course, it can be any triple system and any combination
of bodies denoted by indices $(i, j\ne i, k\ne i)$. Ergo,
$m^\star$~denotes the mass of the Moon,
$m'$~mass of the test particle,
$R$~radius of the Earth,
$k_2$~the Love number of the Earth,
$\Delta t$~time lag,
$\vec r$~vector Earth--Moon (i.e., perturbing body),
$\vec v$~orbital velocity of the Moon,
$\vec r'$~vector Earth--test particle (interacting body),
$\oldvec\omega$~spin rate vector of the Earth,
$\oldvec\Gamma$~torque acting on Earth's spin.
This general formula is used to compute cross-tides among all triples.
In our case, non-negligible interactions are expected for
Kleopatra--1st moon--1st moon,
Kleopatra--2nd moon--2nd moon;
where the tidal dissipation occurs in Kleopatra itself.
Both moons have to be accounted for,
because they contribute to the total torque (spin-down).
A simple Euler integrator is then used to evolve spins,
assuming principal-axis rotation.
The time steps were $0.02\,{\rm d}$ (orbital) 
and $1\,{\rm d}$ (spin). 

\begin{figure}
\centering
\includegraphics[width=9cm]{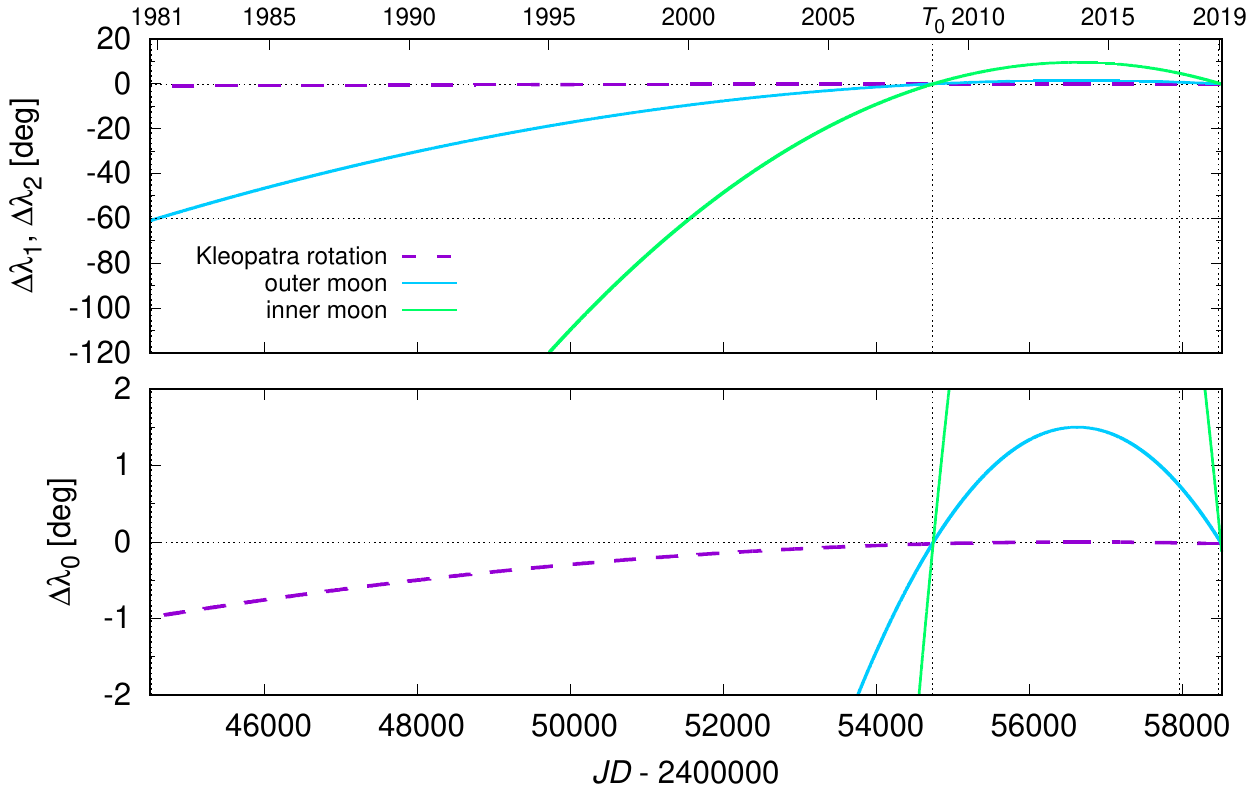}
\caption{Tidal evolution of Kleopatra spin (dashed magenta)
and moon orbits (solid green, blue),
computed as a difference of the true longitudes
$\Delta\lambda_0$, $\Delta\lambda_1$, $\Delta\lambda_2$
between dynamical models with and without tides.
The value of the time lag $\Delta t = 47\,{\rm s}$
corresponds to $\dot P_2$ in Fig.~\ref{map_P2_fitphase0_1.8e-8_ALL}.
The epoch when mean periods coincide was arbitrarily shifted ($\leftrightarrow$)
towards 2456500. Moreover, the mean periods were adjusted ($\updownarrow$)
to fit observations in 2008 and 2018.
}
\label{TEST_Kleopatra_BOTH_spin_E}
\end{figure}

There are three relevant radii of Kleopatra:
$R = 59.6\,{\rm km}$ (volume-equivalent),
$69.0\,{\rm km}$ (surface-equivalent), and
$135\,{\rm km}$ (maximal).
The volume equivalent is commonly used,
but if tidal dissipation happens in surface layers,
the surface equivalent should be preferred.
In case of Kleopatra, we decided to use the maximal radius,
because the strongest dissipation is expected
at the `extremes' of the elongated body.
Other parameters are the Love number
$k_2 = 0.305$
(here, we used the same value as for the Earth),
and the moment of inertia
$I = 1.72\cdot10^{28}\,{\rm kg}\,{\rm m}^2$,
as derived from the ADAM model \citep{Marchis_2021A&A...653A..57M}.
We varied only the time lag and obtained
$\Delta t = 47\,{\rm s}$,
so that the offset in true longitude is
$\Delta\lambda_2 = -60^\circ$
with respect to the model without tides,
or ${\sim}\,0^\circ$ with respect to the observation (occultation).
The evolution is shown in Fig.~\ref{TEST_Kleopatra_BOTH_spin_E}.
It is so smooth because we included only the monopole for Kleopatra
and we overplotted orbits computed separately, without perturbations.

For comparison, the inner (1st) moon should tidally evolve with
$\dot P_1 = 5.0\cdot10^{-8}$,
which is inevitably larger than $\dot P_2 = 1.8\cdot10^{-8}$
due to smaller distance.
The accumulated change in the rotation phase of Kleopatra
due to both moons over the entire time span 1980--2018
then should reach $1^\circ$ (see Sec.~\ref{sec:P0}).

\subsection{Multipole model including tides}

In order to have a complete dynamical model,
we also implemented tides (Eqs.~(\ref{eq:f_tides})--(\ref{eq:Gamma}))
in \xitau\footnote{\url{http://sirrah.troja.mff.cuni.cz/~mira/xitau/}}
\citep{Broz_2017ApJS..230...19B,Broz_2021A&A...653A..56B},
which enabled us to fit all observations.
Let us recall that the model already included
multipoles up to the order $\ell = 10$,
mutual moon perturbations,
and that our previous best-fit model
(\citealt{Broz_2021A&A...653A..56B}; without the 1980 occultation) had $\chi^2 = 368$.%
\footnote{More specifically,
the individual contributions were
$\chi^2_{\rm sky} = 113$ (absolute astrometry),
$\chi^2_{\rm sky2} = 66$ (relative astrometry),
$\chi^2_{\rm ao} = 621$ (adaptive-optics),
and the joint metric was given as
$\chi^2_{\rm sky} + \chi^2_{\rm sky2} + 0.3\chi^2_{\rm ao}$.}

\begin{figure}
\centering
\includegraphics[width=9cm]{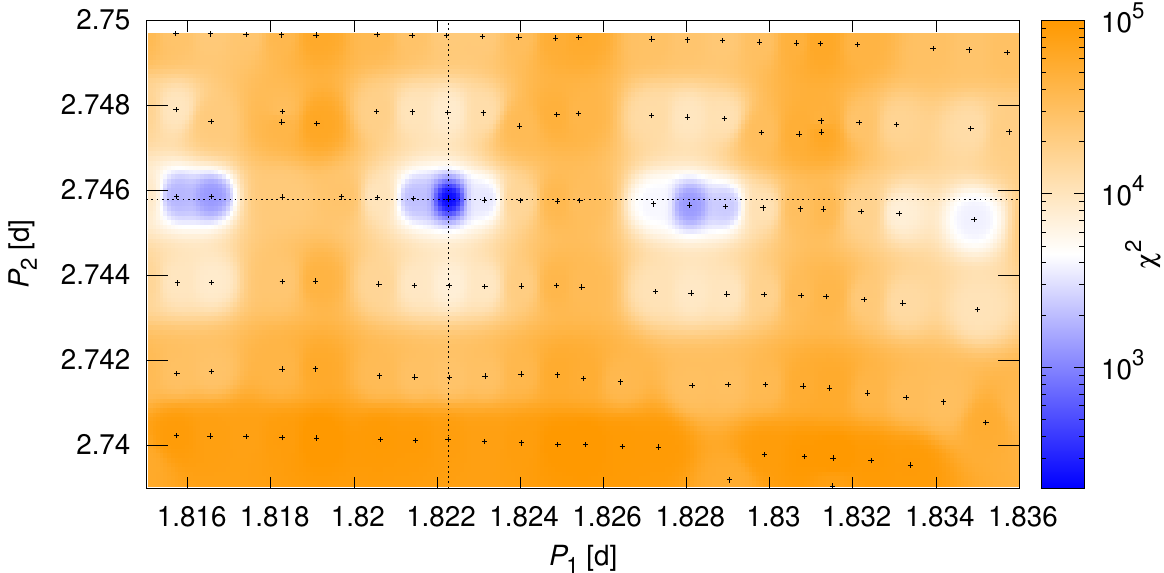}
\caption{$\chi^2 = \chi^2_{\rm sky} + \chi^2_{\rm sky2}$ values for a range
of osculating periods~$P_1$, $P_2$ and converged models.
All black crosses correspond to local minima of $\chi^2$;
colours are interpolated.
A normal $\chi^2$ map would be much more irregular.
The dotted lines show the periods of the global minimum.}
\label{216_fitting35_MIGNARDGRID_P1_P2_min}
\end{figure}

\begin{figure}
\centering
\begin{tabular}{c}
\kern.5cm including tides \\
\includegraphics[width=8.5cm]{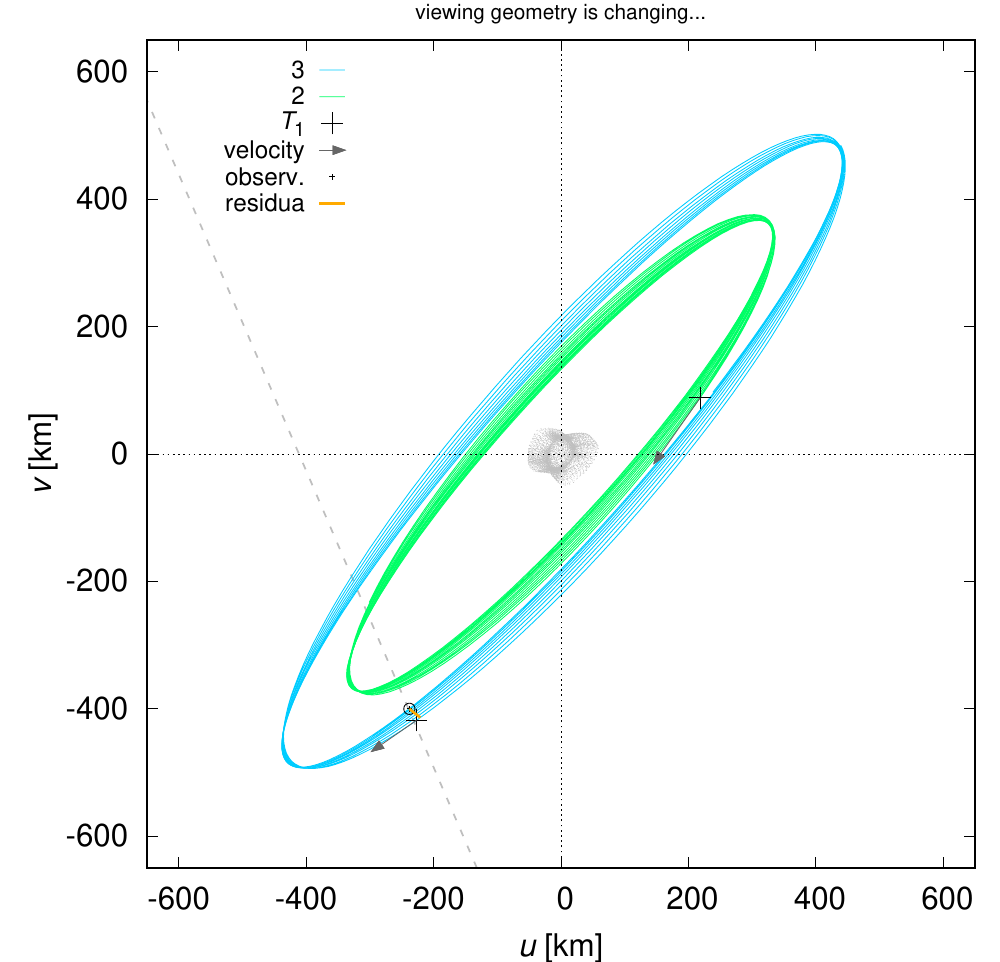} \\
\end{tabular}
\caption{Same as Fig.~\ref{216_test57_SKYANDTELESCOPE_chi2_SKY_uv},
but for our new multipole model including tides.
The offset in true longitude~$\lambda_2$ is negligible
(comparable to the uncertainty).
}
\label{216_fitting36_MIGNARDQ_SKYTEL_chi2_SKY_uv}
\end{figure}

We proceeded in several steps:
(i)~we unsuccessfully tried to re-converge periods $P_1$, $P_2$
(without tides), but the value of $\chi^2$ remained too high, $\chi^2 = 677$,
compared to the number of measurements (reported in Tab.~\ref{tab4});
(ii)~we successfully converged $P_1$, $P_2$
together with a non-zero time lag $\Delta t$ and obtained $\chi^2 = 388$;
(iii)~we verified there is no deeper local minimum in the surroundings
(see Fig.~\ref{216_fitting35_MIGNARDGRID_P1_P2_min});
(iv)~we converged all remaining parameters,
with the final $\chi^2 = 360$
(see Fig.~\ref{216_fitting36_MIGNARDQ_SKYTEL_chi2_SKY_uv}).
The respective parameters are presented in Tab.~\ref{tab4}.

Although multipole perturbations ($4\,{\rm km}$ in $a_2$)
or mutual perturbations ($2\,{\rm km}$)
are orders of magnitude larger than tides ($1\,{\rm m}\,{\rm y}^{-1}$ in $\dot a_2$),
the former are strictly conservative/periodic
and the latter dissipative/non-periodic.
Tides are crucial to explain the 1980 occultation.

\begin{figure}
\centering
\begin{tabular}{c@{\hskip.1cm}c}
& \kern.4cm SPHERE2017 \\
\noalign{\vskip5pt}
\raise3.0cm\hbox{\rotatebox{90}{no tides}} &
\includegraphics[width=7.5cm]{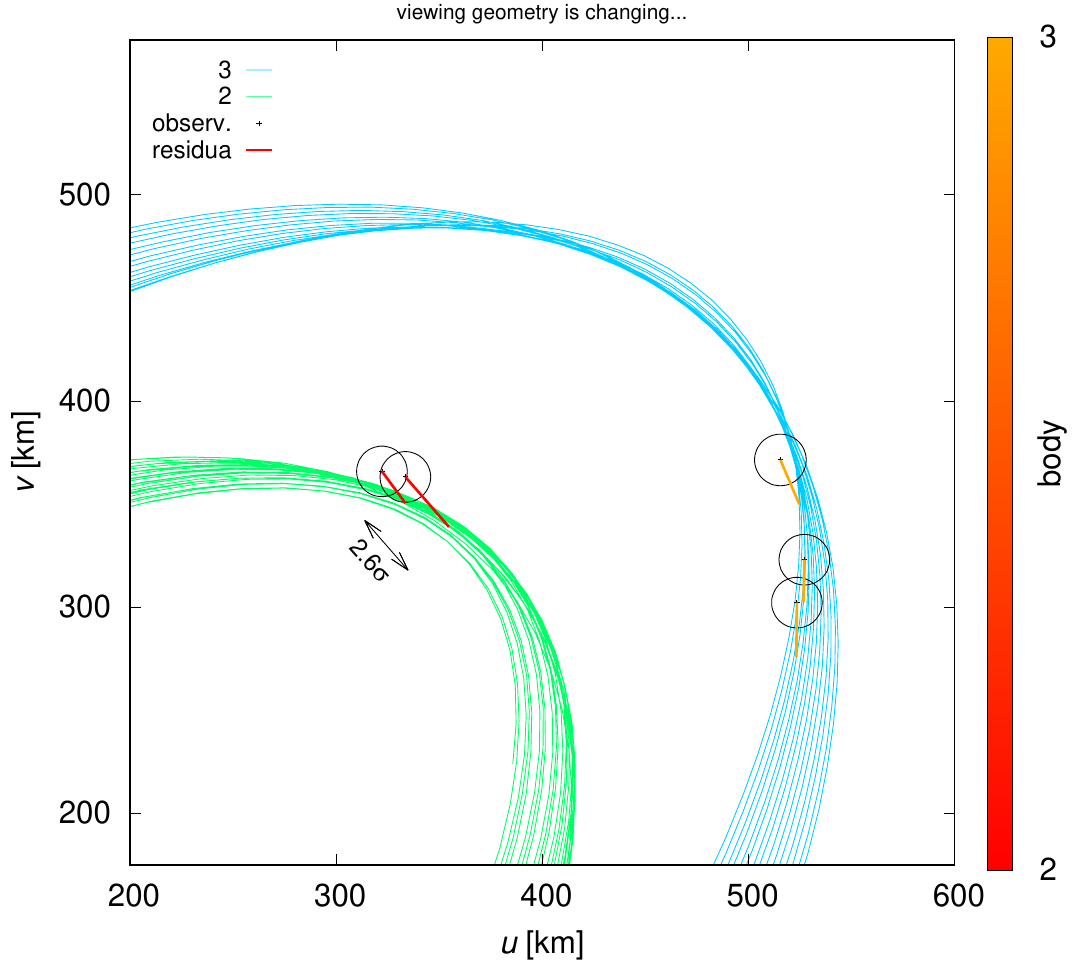} \\
\raise2.6cm\hbox{\rotatebox{90}{including tides}} &
\includegraphics[width=7.5cm]{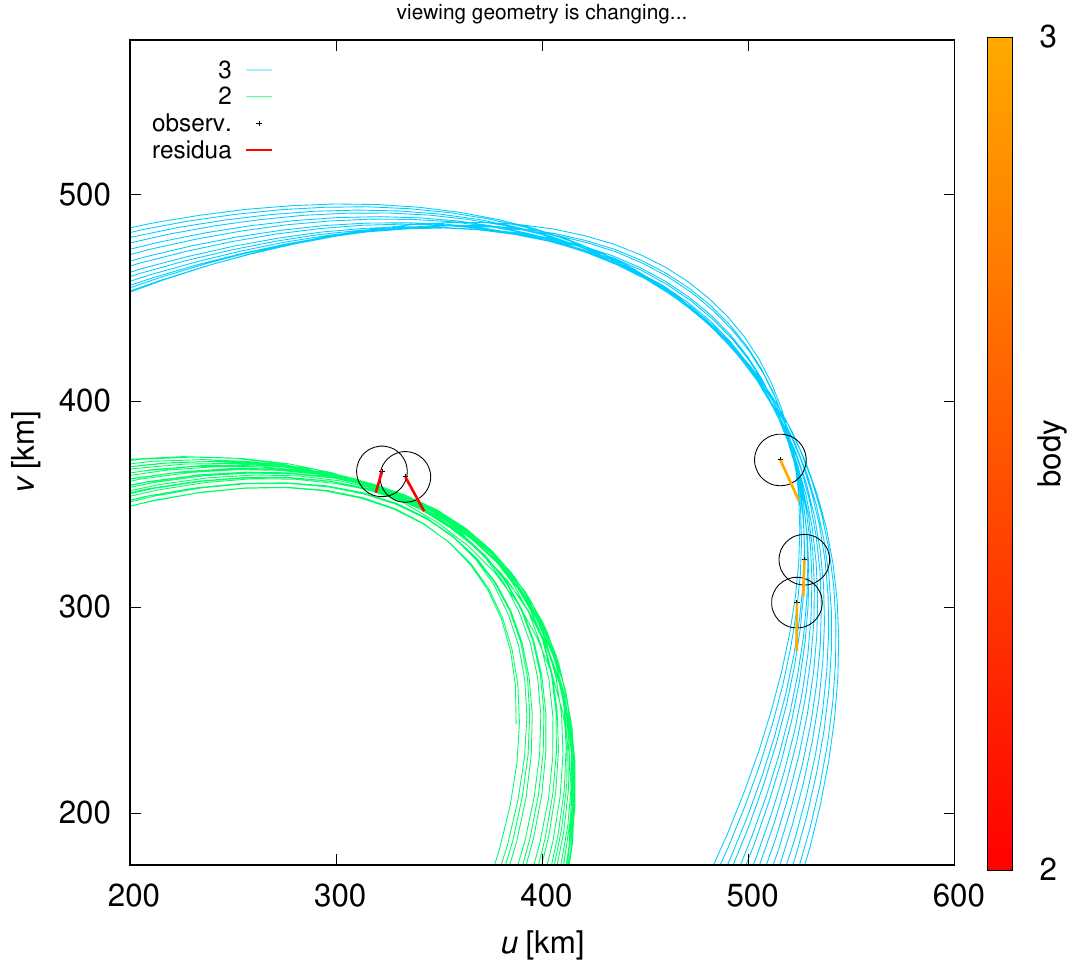} \\
\end{tabular}
\caption{Detail of some SPHERE2017 astrometric observations
and converged models with no tides (top) and including tides (bottom).
The assumed uncertainties ($10\,{\rm mas}$) are indicated by black circles,
residua by red or orange lines.
There is a noticeable improvement for the 1st moon.
However, the 2nd moon is still offset,
possibly due to some remaining systematics.
The proper motion in the $(u,v)$ plane is relatively slow
due to the orbit orientation and the line of sight.
}
\label{216_test57_SKYANDTELESCOPE_SPHERE2017_chi2_SKY_uv}
\end{figure}

Moreover, the tidal evolution may partially explain systematics
in our previous fitting of SPHERE2017 dataset. When the osculating
periods are constant and constrained by DESCAMPS and SPHERE2018,
there have to be some offsets (of the order of $10\,{\rm mas}$)
for the intermediate dataset, especially for the 1st moon
which is more affected by tides. A detailed comparison shows
that the offsets may be decreased when tides are included
(see Fig.~\ref{216_test57_SKYANDTELESCOPE_SPHERE2017_chi2_SKY_uv}).
However, tidal evolution cannot explain {\em all\/} remaining systematics
(cf. our discussion of astrometry in \citealt{Broz_2021A&A...653A..56B}).

\subsection{Possibly increasing rotation period $P_0$}\label{sec:P0}

As we discussed in Sect.~\ref{sec:monopole}, if the moons are affected by tides, so must be the rotation of Kleopatra. If the period is evolving in time, then the value of $P_0 = 5.3852824(10)\,{\rm h}$ reported in \cite{Marchis_2021A&A...653A..57M} corresponds to the middle of the 1977--2018 time span. To estimate a realistic uncertainty of this `mean' rotation period, we created 1000 bootstrapped samples of the light curve data set (random selection of light curves and random selection of points in those light curves) and used them as input for convex light curve inversion. The data set of \cite{Marchis_2021A&A...653A..57M} was supplemented with other observations that are listed in Table.~\ref{tab:new_lcs}. Now it consists of 198 light curves covering the interval 1977--2021. This led to the mean rotation period of:
\begin{equation}
\bar P_0 = (5.3852827\pm0.0000003)\,{\rm h}\,.
\end{equation}
This improved uncertainty of the rotation period corresponds to uncertainty in Kleopatra's rotation phase of $1.3^\circ$ over the interval of 44 years, which is of the same order as the expected $1^\circ$ shift estimated in Sect.~\ref{sec:monopole}.

To check if the predicted deceleration of the main body's rotation is 'visible' in the data, we divided light curves into two sets: the first one covering the interval 1977--1994 and the second one 2002--2021. If the rotation period is changing, we should see some difference in the periods for these two data sets. Similarly like with the full data set, we created 1000 bootstrapped samples and performed the light curve inversion independently for all of them to estimate parameter errors. For the interval 1977--1994, the rotation period was:
\begin{equation}
\bar P_0^{1977\mbox{--}1994} = (5.3852821\pm0.0000010)\,{\rm h}
\end{equation}
and the corresponding phase shift $1.8^\circ$. For 2002--2021, the values were:
\begin{equation}
\bar P_0^{2002\mbox{--}2021} = (5.3852822\pm0.0000005)\,{\rm h}
\end{equation}
and $1.0^\circ$. So the uncertainty intervals are larger (due to shorted time base) than with the full data set and they overlap, i.e., there is no indication that the rotation period is changing. Controversially, the mean period derived from 1977--2021 observations is a bit longer than periods for 1977--1994 and 2002--2021 subsets, while we would expect it to be somewhere in between the two values. This is partly caused by the correlation between the period and pole direction (that is also optimized for each bootstrapped sample) but we think that the main reason are some small but systematic errors present in some light curves. 

To test the sensitivity of our approach, we generated an equivalent set of synthetic observations using the non-convex ADAM shape model from \cite{Marchis_2021A&A...653A..57M}, Hapke's light scattering model, and two values of $\dot P_0$, $3.2 \cdot 10^{-12}$ and $1.6 \cdot 10^{-12}$. We then treated the synthetic data set as real data and applied the same bootstrap approach to detect possible changes in rotation period. For $\dot P_0 \simeq 3.2 \cdot 10^{-12}$, the effect of changing period was clearly visible as a systematic difference between periods for 1977--1994 and 2002--2021 data. This way we checked that convex/non-convex models do not affect the results in a systematic way. However, when using $\dot P_0 \simeq 1.6 \cdot 10^{-12}$ and adding $2\%$ random noise to our synthetic light curves (which is a realistic estimate of observational uncertainties), the effect of changing period was no more detectable --- both subsets of bootstrapped light curves had statistically the same rotation period.

\begin{figure}
\centering
\includegraphics[width=8.5cm]{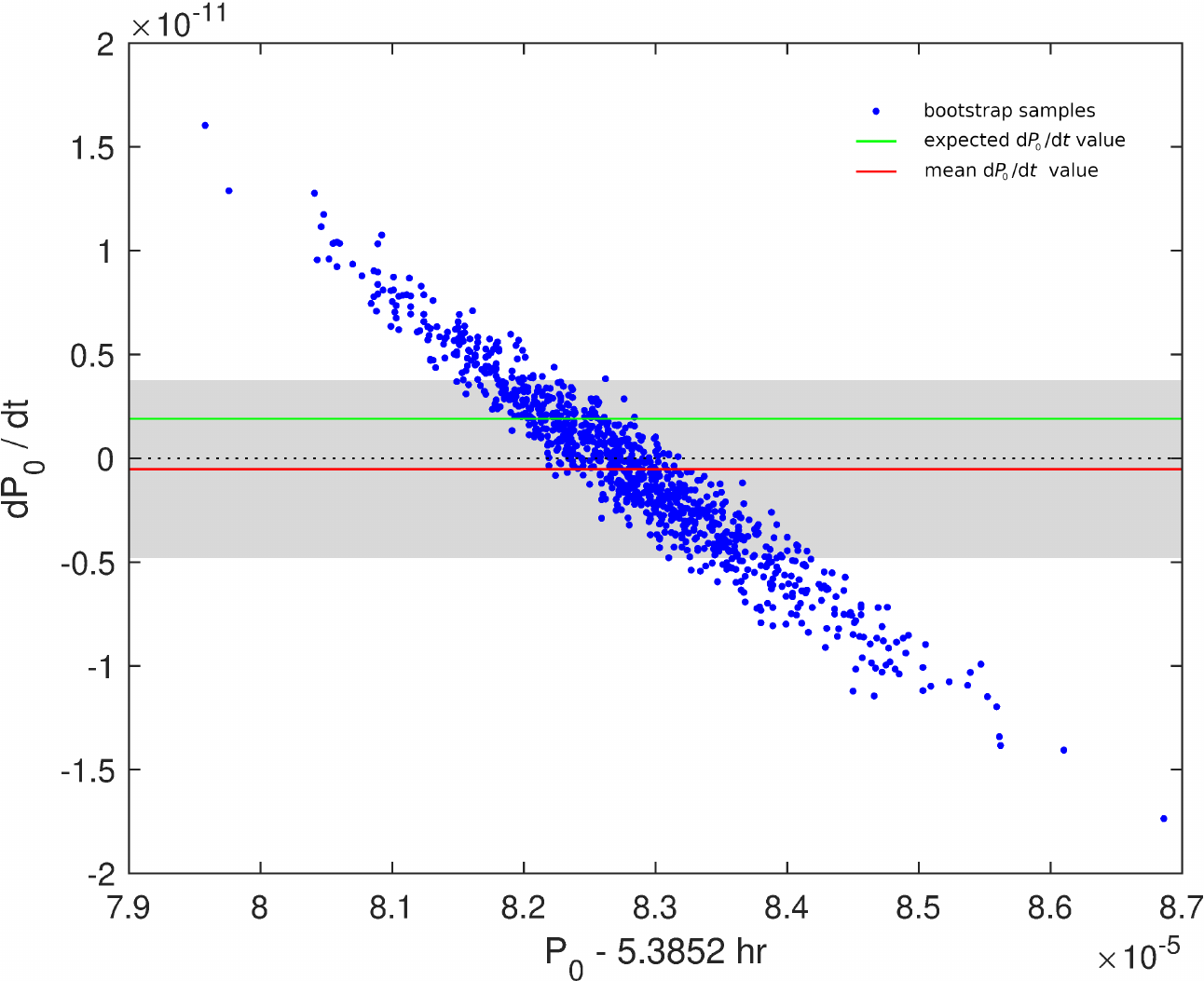}
\caption{Period $P_0$ and its change $\dot P_0$ for 1000 bootstrap samples of the photometric data set. Each blue point represents one bootstrap run. The mean value $-0.5\cdot 10^{-12}$ of $\dot P_0$ is marked with red line, the theoretical prediction $1.9\cdot 10^{-12}$ of Kleopatra's deceleration due to tides is marked with the green line. The gray strip marks 1-$\sigma$ uncertainty interval for $\dot P$.}
\label{fig:dotP_Kleopatra}
\end{figure}

We also tried to detect a possible evolution of Kleopatra's rotation period by including $\dot P_0$ as a {\em free\/} parameter into light curve inversion. In practice, we used the same approach as \cite{Kaa.ea:07} or \cite{Dur.ea:18a} when searching for the YORP effect that influences light curves the same way --- rotation period changes linearly in time (more exactly, angular velocity changes linearly in time but the difference is negligible). We used the same bootstrap sample as in case of fitting light curves with a constant-period model. The results are shown in Fig.~\ref{fig:dotP_Kleopatra}, where $P_0$ is plotted against $\dot P_0$. There is a strong anticorrelation between these two parameters --- positive $\dot P_0$ (deceleration of the rotation) and shorter initial rotation (at the beginning of the observing time interval in 1977) has a similar outcome as negative $\dot P_0$ (acceleration of the rotation) and slower initial rotation. From bootstrap, $\dot P = (-0.5 \pm 4.2)\cdot 10^{-12}$, which means that the effect we are searching for $\dot P_0 = 1.9 \cdot 10^{-12}$ is consistent with the data but cannot be confirmed. Zero $\dot P_0$ is also compatible with the data. Due to correlation, the marginal uncertainty of $P_0$ is 0.0000009\,h, which is larger than when assuming $\dot P_0 = 0$.

\subsection{Discussion of the quality factor~$Q$}\label{sec:Q}

\begin{figure}
\centering
\includegraphics[width=9cm]{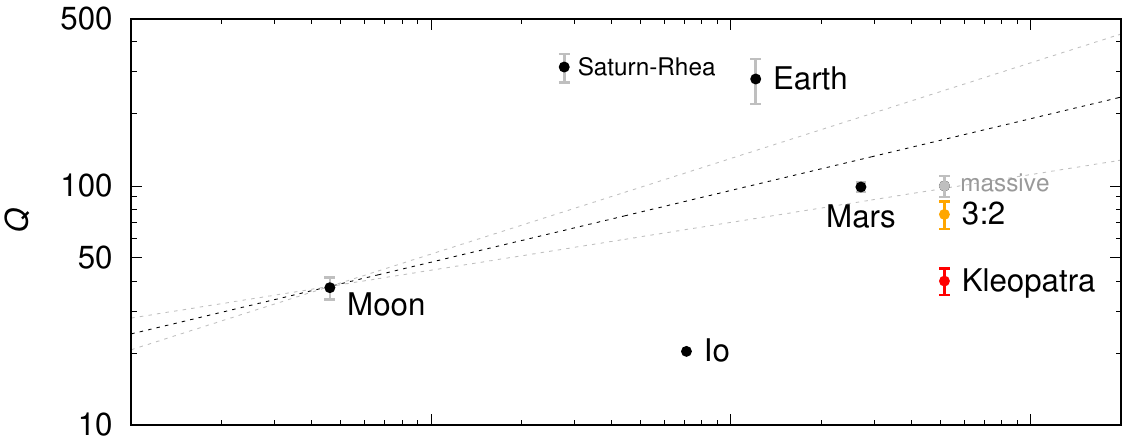}
\includegraphics[width=9cm]{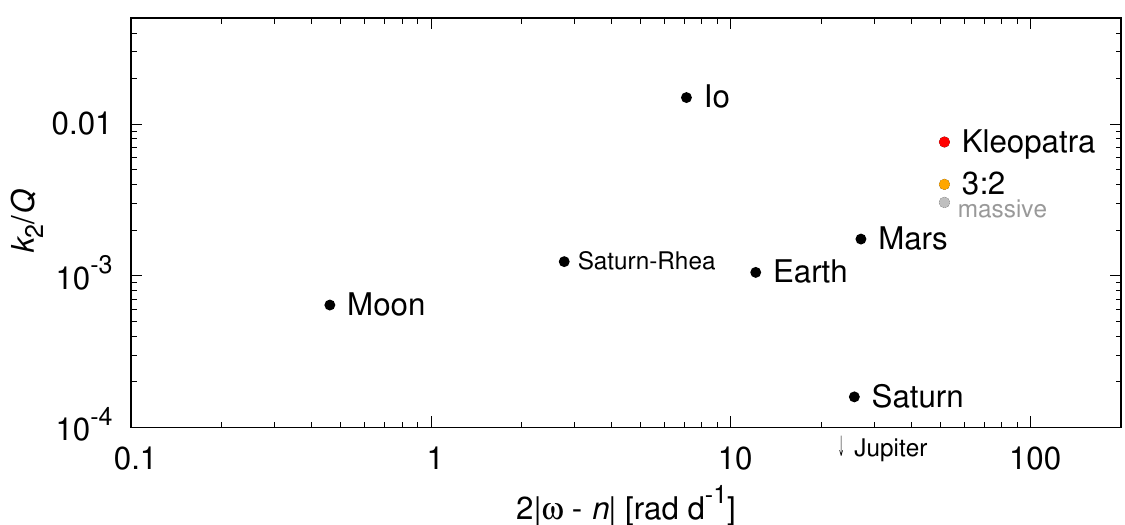}
\caption{Top: Comparison of the quality factors~$Q$ for terrestrial bodies
and Kleopatra, which experience different loading frequencies $\xi \equiv 2|\omega-n|$.
Data from \cite{Lainey_2016CeMDA.126..145L}.
For Io and Kleopatra, $Q$ was estimated from $Q/k_2$ and $k_2 \doteq 0.3$.
The value denoted `3:2' was derived for Kleopatra
when its moons are locked in the 3:2 resonance (see Sec.~\ref{sec:3:2}).
Similarly, `massive' is for the model with more massive moons
(see Sec.~\ref{sec:massive}).
The dotted line is the expected dependence of $Q(\xi)$
(normalized with respect to the Moon;
\citealt{Efroimsky_2007JGRE..11212003E}).
Bottom: Comparison of the ratios $k_2/Q$ vs. $\xi$,
which are directly constrained by the respective tidal evolution.
Kleopatra's is slightly above terrestrial bodies.
}
\label{Q}
\end{figure}

Our modelling of tidal evolution indicates the time lag around
$\Delta t = 42\,{\rm s}$,
with the assumed value of the Love number $k_2 \doteq 0.3$.
According to the approximate relation \citep{Efroimsky_2007JGRE..11212003E}:
\begin{equation}
Q = {1\over\Delta t\,2|\omega_0-n_2|}\,,
\end{equation}
where
$Q$~denotes the quality factor,
$\omega_0 \equiv 2\pi/P_0$ spin rate,
$n_2$ mean motion,
it corresponds to $Q = 40$,
or $Q/k_2 = 131$.
This~$Q$ value is relatively low (i.e., dissipation high),
which seems reasonable for (216) Kleopatra --- an irregular body
close to the critical rotation \citep{Marchis_2021A&A...653A..57M}.
The value of $k_2$ can be hardly orders-of-magnitude lower,
because $Q$ would be unrealistically low.
For comparison, the Earth and Moon have
$Q = 280\pm60$ and $38\pm4$, respectively
\citep{Konopliv_2013JGRE..118.1415K,Lainey_2016CeMDA.126..145L},
but they correspond to low loading frequencies,
$\xi \equiv 2|\omega - n|$,
and the expected dependence $Q(\xi)$ is positive
($Q \propto \xi^{0.3}$ for $\xi \gtrsim 10^{-2}\,{\rm rad}\,{\rm d}^{-1}$;
\citealt{Efroimsky_2007JGRE..11212003E}).
This is demonstrated in Fig.~\ref{Q}.

%
%
%
%
%

For uniform bodies, there is a relation between
the Love number~$k_2$ and the material rigidity~$\mu$
(\citealt{Goldreich_2009ApJ...691...54G}; Eq.~(24)):
\begin{equation}
\mu = \left({3\over 2k_2}-1\right){6\over 19}{Gm_1^2\over R^2}{1\over S} \doteq {9\over 19}{1\over k_2}{Gm_1^2\over R^2}{1\over S}\,,
\end{equation}
where $S$ denotes the surface area;
the approximation holds for bodies with substantial~$\mu$
(or small~$k_2$).
Because we know $Q/k_2$, it allows us to obtain
$\mu Q = 2.7\cdot10^7\,{\rm Pa}$.
This is the same order of magnitude as the estimate
for 1-km asteroids \citep{Scheirich_2015Icar..245...56S},
but three orders of magnitude smaller than the value
$\mu Q \simeq 10^{10}\,{\rm Pa}$
derived for other 100-km asteroids \citep{Marchis_2008Icar..196...97M,Marchis_2008Icar..195..295M}.
We can also try to express $\mu = 6.7\cdot 10^5\,{\rm Pa}$
(from $k_2$), which is not independently constrained, though.
It seems compatible with loose material,
or at least regolith-covered bodies.


There is also a relation to the regolith thickness
(\citealt{Nimmo_2019Icar..321..715N}; Eq.~(6)):
\begin{equation}
l = \sqrt{m_3 n_2^2 R^2\over 3 m_1 G\rho f Q}\,,
\end{equation}
where $f = 0.6$ is the assumed friction coefficient,
and resulting $l = 13\,{\rm m}$.
Of course, for non-spherical bodies, there may be significant deviations.
In particular, when we used the maximum radius~$R$
and only a part of the surface is at this distance,
the regolith needed to explain all the dissipation
is probably accordingly thicker.


\subsection{$Q$ for orbits in the 3:2 resonance}\label{sec:3:2}

The orbits of the two moons appear to be very close to the 3:2 mean-motion resonance; the respective critical angle~$\sigma$ does not librate though,
because orbits are so perturbed by the multipoles of Kleopatra 
and eccentricities are too small \citep{Broz_2021A&A...653A..56B}.
Nevertheless, if they are locked, tides act on both moons at the same time
and, inevitably, $\dot P_2 = 1.5\dot P_1$.
According to our numerical experiments
(using the machinery of Sec.~\ref{sec:monopole}),
the value of $\dot P_1$ decreases,
and $\dot P_2$ increases,
compared to their nominal values.
In order to obtain the same offset of $\Delta\lambda_2 = +60^\circ$,
the required values are now
$\dot P_0 = 0.9\cdot 10^{-12}$,
$\dot P_1 = 1.2\cdot 10^{-8}$,
$\dot P_2 = 1.8\cdot 10^{-8}$.
It corresponds to the time lag of approximately
$\Delta t_1 = 22\,{\rm s}$.

Consequently, the dissipation factor as well as other derived quantities
from Sec.~\ref{sec:Q} are revised as follows:
$Q = 76$,
$Q/k_2 = 250$,
$\mu Q = 5.0\cdot 10^7\,{\rm Pa}$, and
$l = 9\,{\rm m}$.
The assumption of the 3:2 resonance thus decreases the dissipation rate
and puts Kleopatra somewhat closer to the theoretical dependence
of $Q(\xi)$ on Fig.~\ref{Q}.



\subsection{Q for more massive moons}\label{sec:massive}

In an alternative model, moons can be more massive (more dense
than Kleopatra), with
$m_2 = 4\cdot 10^{-16}\,M_{\rm S}$,
$m_3 = 9\cdot 10^{-16}\,M_{\rm S}$
\citep{Broz_2021A&A...653A..56B},
and the deformation potential is proportionally larger
(Eq.~(\ref{K1})).
Again, to obtain $\Delta\lambda_2 = +60^\circ$,
$\Delta t_2 = 16\,{\rm s}$ is required,
together with 
$\dot P_0 = 3.6\cdot 10^{-12}$,
$\dot P_1 = 3.1\cdot 10^{-8}$,
$\dot P_2 = 1.8\cdot 10^{-8}$.
The value of $P_0$ is increased substantially,
but still not enough to be confirmed (or excluded) by observations.
Adjustments of other parameters are as follows:
$Q = 100$,
$Q/k_2 = 330$,
$\mu Q = 8.2\cdot 10^7\,{\rm Pa}$, and
$l = 13\,{\rm m}$.
This puts Kleopatra even closer to the theoretical dependence
on Fig.~\ref{Q} and indicates that mechanical properties
of Kleopatra's material may actually be similar to the terrestrial bodies. 



\subsection{Discussion of the origin}

\begin{figure}
\centering
\includegraphics[width=9cm]{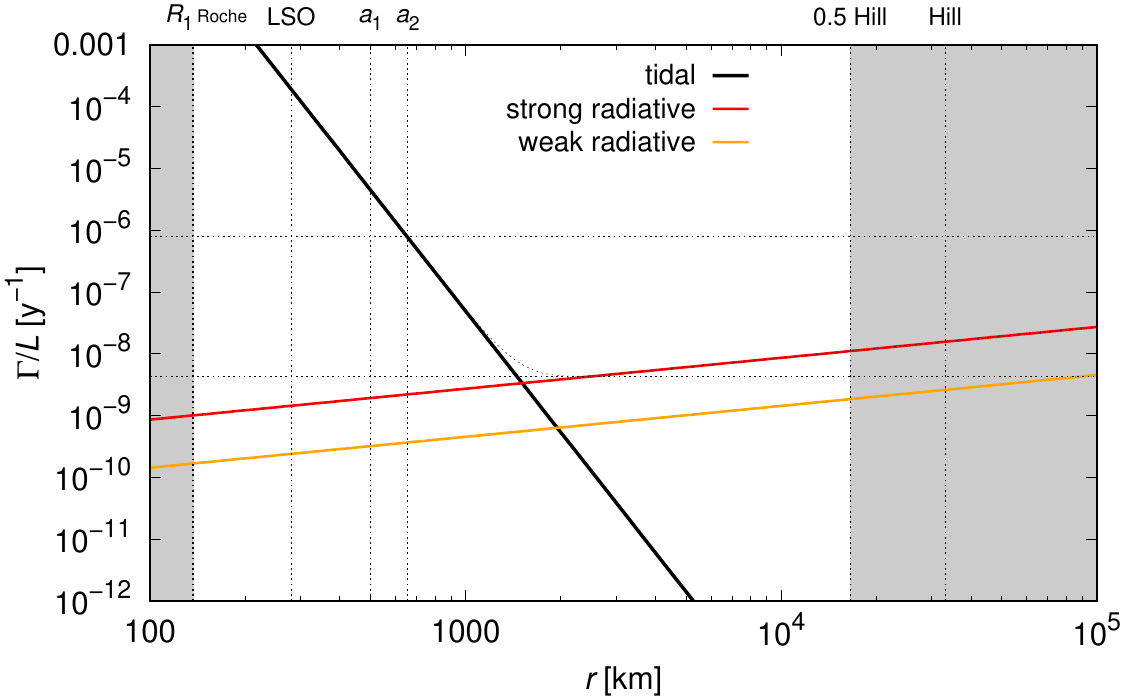}
\caption{Torque~$\Gamma$ over angular momentum $L$
(in ${\rm y}^{-1}$ units) for the tidal (black),
strong radiative (red),
and weak radiative torques (orange).
Relevant radial~$r$ are indicated (vertical dotted lines):
the maximum radius~$R_1$ of the primary,
Roche radius~$r_{\rm R} = R_1(2\rho_2/\rho_0)^{1/3}$,
last stable orbit (LSO),
semimajor axes~$a_1$, $a_2$ of the moons,
half of $r_{\rm H}$,
and the Hill radius~$r_{\rm H} = a_{\rm h}(m_1/(3M_\odot))^{1/3}$.
The corotation orbit does not exist.
For (216) Kleopatra, tidal and radiative torques $|\Gamma|$
become comparable at $r = 1500$ to $2000\,{\rm km}$.
}
\label{vypocty20_BYORP}
\end{figure}

Regarding the origin of the moons, it is interesting to estimate
the time scale as the angular momentum over the tidal torque,
$L_2/\Gamma_2 \simeq 1.3\cdot 10^6\,{\rm y}$,
because it would indicate the moons are very young.
The dependence of both tidal and radiative torques,
computed for the Kleopatra system according to
Eqs.~(\ref{Gamma_L_radiative}), (\ref{Gamma_L_tidal}),
is shown in Fig.~\ref{vypocty20_BYORP}.
If the initial distance coincided with the last stable orbit,
at about $r_{\rm lso} = 280\,{\rm km}$
(or $P \simeq 0.8\,{\rm d}$)
according to our numerical tests,
and the final distance is comparable to half of the Hill sphere,
$r_{\rm H} = 33100\,{\rm km}$,
the overall evolution would take over
$2\cdot 10^{8}\,{\rm y}$.%
\footnote{assuming the BYORP would not be interrupted by periods
of non-synchronous rotation of the moon}
In a broader perspective, this is comparable to
the dynamical time scale of Saturn rings
(\citealt{Charnoz_2009sfch.book..537C};
although cf. \citealt{Crida_2019NatAs...3..967C}).

The moons are definitely younger than Kleopatra, because a large-scale
collisional event would leave observable traces (an asteroid family).
The moons may be rather related to small-scale craterings,
which are much more frequent. Out of three options:
(i)~a cratering with a direct reaccretion of multi-kilometre moons;
(ii)~a collisional spin-up of Kleopatra over its critical frequency and mass shedding;
(iii)~low-speed ejection of material from the surface below the L1 critical point
(see Fig.~6 in \citealt{Marchis_2021A&A...653A..57M}) and continuous accretion from ring;
the last one seems the easiest.

However, the long-term evolution could be complicated.
If Kleopatra has been close to its rotation limit
for a prolonged period of time, moons have been created often.
It implies there are perhaps more moons within the Hill sphere,
as suggested on some of Keck images.
The most likely distance seems to be about 1500\,km,
where $\Gamma/L$ is lowest and evolution slowest.
Such a hypothetical 3rd moon would be close to the 3:1 resonance
with the 2nd moon and a capture is inevitable.
Subsequent evolution of eccentricity, which is increasing by tides
\citep{Goldreich_1963MNRAS.126..257G,Correia_2012ApJ...744L..23C},
would lead to an instability of the moon system and an ejection
of one or two moons beyond the Hill sphere.
The time scale of evolution is determined by the inner moon.
The instability may be delayed by the protective resonant mechanism,
or alleviated if the moons has been rotating synchronously
(1:1) and dissipating due to higher tidal modes (3:2, 2:1).


\section{Conclusions}


Hereinafter, we summarise that astrometric and occultation observations of Kleopatra's outer moon
indicate a secular evolution of its orbital period
$\dot P_2 = (1.8\pm0.1)\cdot 10^{-8}$,
which is the first such observation in a system of moons orbiting a large (100-km) asteroid.
It should be linked to a secular evolution of the rotation period
$\dot P_0 = 1.9\cdot 10^{-12}$
of (216) Kleopatra itself.
The latter value is not excluded by current photometric observations,
but their precision (about $1^\circ$ in phase, or $3\,$miliseconds in period)
is still not sufficient to exclude $\dot P_0 = 0$.

For future observers, we predict a secular evolution of the 1st moon
$\dot P_1 \simeq 5.0\cdot 10^{-8}$, which is inevitable when the 2nd moon
is driven by tides. If the observed value will be different,
it could indicate, e.g.,
stronger mutual interactions,
different masses~$m_2$, $m_3$ of the moons,
or a greater proximity to the 3:2 mean-motion resonance.
If the moons are inside the 3:2 resonance,
the tides acting on the 1st moon
also act on the 2nd moon,
and a lower dissipation in Kleopatra is sufficient
to explain the offset in true longitude $\lambda_2$.
In more complex rheological models the time lag $\Delta t$
(or~$Q$) also depends on loading frequencies, i.e., $2|\omega-n|$.
However, in the Kleopatra triple system, the loading frequencies
are perhaps too close ($49.1$, $51.4\,{\rm rad}\,{\rm d}^{-1}$)
to measure this dependence directly, by means of accurate astrometry.

At the same time, adaptive-optics observations
of fast-moving shadows (at higher phase angles)
can be perhaps used to better constrain the rotation phase of Kleopatra
and detect a possible difference between measured $\dot P_0$
and $\dot P_0'$ inferred from tides
(similarly as in the Earth--Moon system; cf.~post-glacial rebound).
Consequently, ground-based observations with the VLT/SPHERE
instrument have a potential to constrain `geophysical' internal evolution
of large asteroids.

\begin{figure}
\includegraphics[width=8cm]{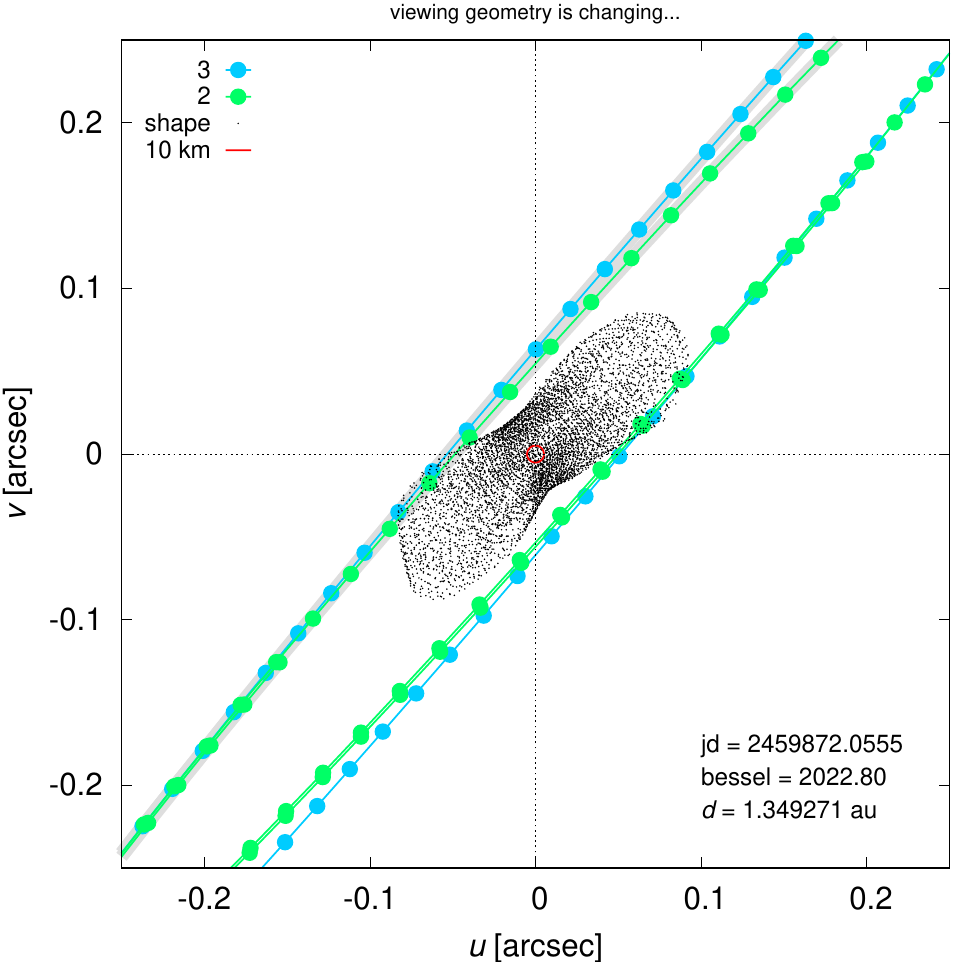}
\caption{Sky-plane projection of Kleopatra and moon orbits
for the Besselian year 2022.80 (Oct),
i.e. one of the epochs when eclipses and transits will be observable.
The spacing between points corresponds to 0.02\,d.
Approximate sizes of the moons are 10\,km,
corresponding to 5\,mas.
}
\label{chi2_SKY_uv_2022.80}
\end{figure}

Another opportunity to observe (216) Kleopatra and its moons will
be in 2022--2024. According to our ephemeris, transits and eclipses
of the moons will occur (e.g., Fig.~\ref{chi2_SKY_uv_2022.80}).
The intervals when orbital planes cross Kleopatra are as follows:
\begin{center}
\begin{tabular}{lll}
2022.34-2022.41 & May       & 2.32\,au \\
2022.80-2022.87 & Oct-Nov   & 1.34\,au \\
2023.93-2024.05 & Dec-Jan   & 1.94\,au \\
2024.51-2024.59 & Jul       & 3.70\,au \\
\end{tabular}
\end{center}
AO and possibly also precise photometric observations
can help to constrain sizes and albedos of the moons.
This is also true for stellar occultations (see App.~\ref{app:B}).
Regarding hypothetical moons separated by 1500\,km or more,
where radiative torques should be dominant,
a deeper survey with the next-generation AO instruments
like VLT/ERIS or Gemini/GPI2 would be useful.

%
%
%

\begin{table*}
\caption{Best-fit models with no tides (left) and including tides (middle),
together with realistic uncertainties of the parameters (right).}
\label{tab4}
\centering
\begin{tabular}{lrrlr}
var. & val. & val. & unit & $\sigma$ \\
\hline
\vrule height10pt width0pt
$m_1         $ & $   1.492735\cdot10^{-12}   $ & $   1.492735\cdot10^{-12}  $ & $M_{\rm S}$ & $ 0.16\cdot10^{-12} $ \\
$m_2         $ & $   2\cdot10^{-16}          $ & $   2\cdot10^{-16}         $ & $M_{\rm S}$ & $ 2\cdot10^{-16}    $ \\
$m_3         $ & $   3\cdot10^{-16}          $ & $   3\cdot10^{-16}         $ & $M_{\rm S}$ & $ 3\cdot10^{-16}    $ \\
$P_1         $ & $   1.822359                $ & $   1.822281               $ & day         & $ 0.004156          $ \\
$\log e_1    $ & $  -3.991                   $ & $  -3.991                  $ & 1           & $ -3$ (i.e. 0.001)    \\
$i_1         $ & $  70.104                   $ & $  70.104                  $ & deg         & $ 1.0               $ \\
$\Omega_1    $ & $ 252.920                   $ & $ 252.920                  $ & deg         & $ 1.0               $ \\
$\varpi_1    $ & $   0.089                   $ & $   0.089                  $ & deg         & $ 10.0              $ \\
$\lambda_1   $ & $  59.665                   $ & $  59.665                  $ & deg         & $ 1.0               $ \\
$P_2         $ & $   2.745820                $ & $   2.745791               $ & day         & $ 0.004820          $ \\
$\log e_2    $ & $  -3.998                   $ & $  -3.998                  $ & 1           & $ -3                $ \\
$i_2         $ & $  70.347                   $ & $  70.347                  $ & deg         & $ 1.0               $ \\
$\Omega_2    $ & $ 252.954                   $ & $ 252.954                  $ & deg         & $ 1.0               $ \\
$\varpi_2    $ & $   1.601                   $ & $   1.601                  $ & deg         & $ 10.0              $ \\
$\lambda_2   $ & $ 108.357                   $ & $ 108.357                  $ & deg         & $ 1.0               $ \\
$l_{\rm pole}$ & $  72.961                   $ & $  72.961                  $ & deg         & $ 1.0               $ \\
$b_{\rm pole}$ & $  19.628                   $ & $  19.628                  $ & deg         & $ 1.0               $ \\
$\Delta t_1  $ & --                            & $  42.1                    $ & s           & $ 1.0               $ \\
\hline
$n_{\rm sky}         $ & $  68  $ & $  68$ \\
$n_{\rm sky2}        $ & $  28  $ & $  28$ \\
$n_{\rm ao}          $ & $3240  $ & $3240$ \\
\hline
$\chi^2_{\rm sky}    $ & $ 617  $ & $ 110$ \\
$\chi^2_{\rm sky2}   $ & $  66  $ & $  60$ \\
$\chi^2_{\rm ao}     $ & $ 621  $ & $ 621$ \\
$\chi^2              $ & $ 872  $ & $ 360$ \\
\hline
$\chi^2_{\rm R\,sky} $ & $ 9.07 $ & $1.62$ \\
$\chi^2_{\rm R\,sky2}$ & $ 2.35 $ & $2.14$ \\
$\chi^2_{\rm R\,ao}  $ & $ 0.19 $ & $0.19$ \\
\end{tabular}
\tablefoot{
The left model does not fit Oct 10th 1980 occultation
(see Fig.~\ref{216_test57_SKYANDTELESCOPE_chi2_SKY_uv});
without this observation, its $\chi^2$ would be 368.
Orbital elements of the moons are osculating,
for the epoch $T_0 = 2454728.761806$, where
$m_1$ denotes the mass of body~1 (i.e.~Kleopatra),
$m_2$ body~2 (1st moon),
$m_3$ body~3 (2nd moon),
$P_1$ the orbital period of the 1st orbit,
$e_1$ eccentricity,
$i_1$ inclination,
$\Omega_1$ longitude of node,
$\varpi_1$ longitude of pericentre,
$\lambda_1$ true longitude,
etc. of the 2nd orbit;
$l_{\rm pole}$ ecliptic longitude of Kleopatra's rotation pole,
$b_{\rm pole}$ ecliptic latitude;
$n$~numbers of observations (SKY, SKY2, AO),
$\chi^2$~values,
$\chi^2_{\rm R} \equiv \chi^2/n$ reduced values.
The angular orbital elements are expressed in the standard stellar reference frame.
If the orbits lie in the equatorial plane of body~1, they fulfil
$i = 90^\circ-b_{\rm pole}$,
$\Omega = 180^\circ+l_{\rm pole}$.
}
\end{table*}




\begin{acknowledgements}
We thank an anonymous referee for comments.
This work has been supported by the Czech Science Foundation through grant
21-11058S (M.~Bro\v z, D.~Vokrouhlick\'y),
20-08218S (J.~\v Durech, J.~Hanu\v s),
and by the Charles University Research program No. UNCE/SCI/023.
This material is partially based upon work supported by the National Science Foundation under Grant No. 1743015.
B.~Carry and P.~Vernazza were supported by CNRS/INSU/PNP.
This work uses optical data from the Courbes de rotation d'ast\' ero\" ides et de com\` etes database (CdR, \url{http://obswww.unige.ch/\textasciitilde behrend/page\_cou.html}).
The data presented herein were obtained partially at the W. M. Keck Observatory, which is operated as a scientific partnership among the California Institute of Technology, the University of California and the National Aeronautics and Space Administration. The Observatory was made possible by the generous financial support of the W. M. Keck Foundation. The authors wish to recognize and acknowledge the very significant cultural role and reverence that the summit of Maunakea has always had within the indigenous Hawaiian community.  We are most fortunate to have the opportunity to conduct observations from this mountain.
\end{acknowledgements}

\bibliographystyle{aa}
\bibliography{references, bibliography_all}

\newpage
\appendix

\section{List of new light curves}

Observational circumstances of new light curves are provided in Tab.~\ref{tab:new_lcs}.

\onecolumn
\begin{longtable}{rlr rrr l l}
\caption{\label{tab:new_lcs}New optical disk-integrated lightcurves of (216) Kleopatra used in this work.}\\
\hline 
\multicolumn{1}{c} {N} & \multicolumn{1}{c} {Epoch} & \multicolumn{1}{c} {$N_p$} & \multicolumn{1}{c} {$\Delta$} & \multicolumn{1}{c} {$r$} & \multicolumn{1}{c} {$\varphi$} & \multicolumn{1}{c} {Filter} & Observers/Reference \\
 &  &  & (AU) & (AU) & (\degr) &  &  \\
\hline\hline
\endfirsthead
\caption{continued.}\\
\hline
\multicolumn{1}{c} {N} & \multicolumn{1}{c} {Epoch} & \multicolumn{1}{c} {$N_p$} & \multicolumn{1}{c} {$\Delta$} & \multicolumn{1}{c} {$r$} & \multicolumn{1}{c} {$\varphi$} & \multicolumn{1}{c} {Filter} & Reference \\
 &  &  & (AU) & (AU) & (\degr) &  &  \\
\hline\hline
\endhead
\hline
\endfoot
\hline
     1	  &  2002-05-15.0   &  31   &  2.45  &  3.45  &  2.0   &  C      &  Christophe Demeautis         \\
     2	  &  2002-05-15.9   &  17   &  2.45  &  3.45  &  2.2   &  C      &  Christophe Demeautis         \\
     3	  &  2002-05-17.0   &  35   &  2.45  &  3.45  &  2.5   &  C      &  Christophe Demeautis         \\
     4	  &  2003-07-19.0   &  39   &  1.68  &  2.64  &  8.9   &  C      &  Claudine Rinner   \\
     5	  &  2004-12-14.1   &  122  &  1.59  &  2.41  &  15.6  &  C      &  Horacio Correia   \\
     6	  &  2004-12-20.1   &  315  &  1.56  &  2.43  &  13.7  &  C      &  Horacio Correia   \\
     7	  &  2010-04-09.9   &  18   &  2.39  &  3.01  &  16.9  &  C      &  Yassine Damerdji, Jean-Pierre Troncin \\
     & & & & & & & Jean Surej, Philippe Bendjoya  \\
     & & & & & & & Davide Ricci, Raoul Behrend  \\
     & & & & & & & Thierry De Gouvenain, Mugane Diet  \\
     & & & & & & & Mathias Marconi, Jean-By Gros  \\
     & & & & & & & Christophe Giordano, Jean-Christophe Flesch  \\
     & & & & & & & Ivan Belokogne, Andrei Belokogne  \\
     & & & & & & & Axel Bazi  \\
     8	  &  2010-04-09.9   &  5    &  2.39  &  3.01  &  16.9  &  C      & Yassine Damerdji, Jean-Pierre Troncin \\
     & & & & & & & Jean Surej, Philippe Bendjoya  \\
     & & & & & & & Davide Ricci, Raoul Behrend  \\
     & & & & & & & Thierry De Gouvenain, Mugane Diet  \\
     & & & & & & & Mathias Marconi, Jean-By Gros  \\
     & & & & & & & Christophe Giordano, Jean-Christophe Flesch  \\
     & & & & & & & Ivan Belokogne, Andrei Belokogne  \\
     & & & & & & & Axel Bazi  \\
     9	  &  2010-04-09.9   &  6    &  2.39  &  3.01  &  16.9  &  C      & Yassine Damerdji, Jean-Pierre Troncin \\
     & & & & & & & Jean Surej, Philippe Bendjoya  \\
     & & & & & & & Davide Ricci, Raoul Behrend  \\
     & & & & & & & Thierry De Gouvenain, Mugane Diet  \\
     & & & & & & & Mathias Marconi, Jean-By Gros  \\
     & & & & & & & Christophe Giordano, Jean-Christophe Flesch  \\
     & & & & & & & Ivan Belokogne, Andrei Belokogne  \\
     & & & & & & & Axel Bazi  \\
     10	  &  2010-04-26.9   &  72   &  2.64  &  3.04  &  18.7  &  C      &  Jacques Montier, Serge Heterier, Raoul Behrend     \\
     11	  &  2010-05-22.9   &  37   &  3.04  &  3.10  &  18.9  &  C      &  Jacques Montier, Jean-Pierre Previt     \\
     12	  &  2015-01-25.1   &  203  &  2.44  &  3.10  &  15.3  &  C      &  Georg Piehler, Alfons Gabel       \\
     13	  &  2015-01-29.1   &  128  &  2.40  &  3.11  &  14.5  &  C      &  Georg Piehler, Alfons Gabel       \\
     14	  &  2015-02-19.0   &  343  &  2.25  &  3.15  &  9.1   &  C      &  Pierre Antonini       \\
     15	  &  2015-02-19.0   &  67   &  2.25  &  3.15  &  9.0   &  C      &  Matthieu Conjat       \\
     16	  &  2015-02-19.1   &  387  &  2.25  &  3.15  &  9.0   &  C      &  Rene Roy   \\
     17	  &  2015-02-23.0   &  183  &  2.24  &  3.16  &  7.9   &  C      &  Federico Manzini      \\
     18	  &  2015-03-06.0   &  310  &  2.21  &  3.18  &  4.9   &  C      &  Nicolas Esseiva, Raoul Behrend    \\
     19	  &  2017-07-16.0   &  154  &  1.72  &  2.68  &  8.6   &  C      &  Nicolas Esseiva, Raoul Behrend    \\
     20	  &  2017-07-16.0   &  9    &  1.72  &  2.68  &  8.6   &  C      &  Nicolas Esseiva, Raoul Behrend    \\
     21	  &  2017-07-16.0   &  9    &  1.72  &  2.68  &  8.6   &  C      &  Nicolas Esseiva, Raoul Behrend    \\
     22	  &  2017-8-30.3    &  74   &  1.75  &  2.56  &  16.2  &  I      &  Kevin Alton, \citet{Alton2009}   \\
     23	  &  2017-8-31.3    &  115  &  1.75  &  2.56  &  16.5  &  I      &  Kevin Alton, \citet{Alton2009}   \\
     24	  &  2017-8-6.2     &  106  &  1.67  &  2.62  &  9.4   &  I      &  Kevin Alton, \citet{Alton2009}   \\
     25	  &  2017-9-10.1    &  120  &  1.81  &  2.53  &  19.0  &  I      &  Kevin Alton, \citet{Alton2009}   \\
     26	  &  2017-9-11.1    &  124  &  1.82  &  2.53  &  19.2  &  I      &  Kevin Alton, \citet{Alton2009}   \\
     27	  &  2017-9-5.1     &  113  &  1.78  &  2.55  &  17.8  &  I      &  Kevin Alton, \citet{Alton2009}   \\
     28	  &  2017-9-7.3     &  83   &  1.79  &  2.54  &  18.3  &  I      &  Kevin Alton, \citet{Alton2009}   \\
     29	  &  2019-1-4.2     &  889  &  1.49  &  2.43  &  8.7   &  R      &  St\'ephane Fauvaud       \\
     30	  &  2019.1-2019.1  &  543  &  1.50  &  2.45  &  8.2   &  V      &  TESS, \citet{Pal2020}  \\
     31	  &  2021-04-12.1   &  49   &  2.68  &  3.37  &  13.8  &  C      &  David Augustin, Raoul Behrend     \\
     32	  &  2021-04-19.3   &  98   &  2.59  &  3.37  &  12.4  &  C      &  David Augustin, Raoul Behrend   \\
   \hline
\end{longtable}
\tablefoot{
     For each lightcurve, the table gives the epoch, the number of individual measurements $N_p$, asteroid's distances to the Earth $\Delta$ and the Sun $r$, phase angle $\varphi$, photometric filter and the observer(s). Majority of the  data is from the Courbes de rotation d'ast\' ero\" ides et de com\` etes database (CdR, \url{http://obswww.unige.ch/\textasciitilde behrend/page\_cou.html}), maintained by Raoul Behrend at Observatoire de Gen\` eve.
    }

\section{Predictions for stellar occultations 2022--2026}\label{app:B}

Predictions of Kleopatra's moons positions
for expected stellar occultations 2022--2026
are plotted in Fig.~\ref{occultations_2026}.

\def\s#1{\includegraphics[width=4.3cm]{{figs3/216_predictions_2026/occultations2/#1}-eps-converted-to.pdf}}%

\begin{figure}
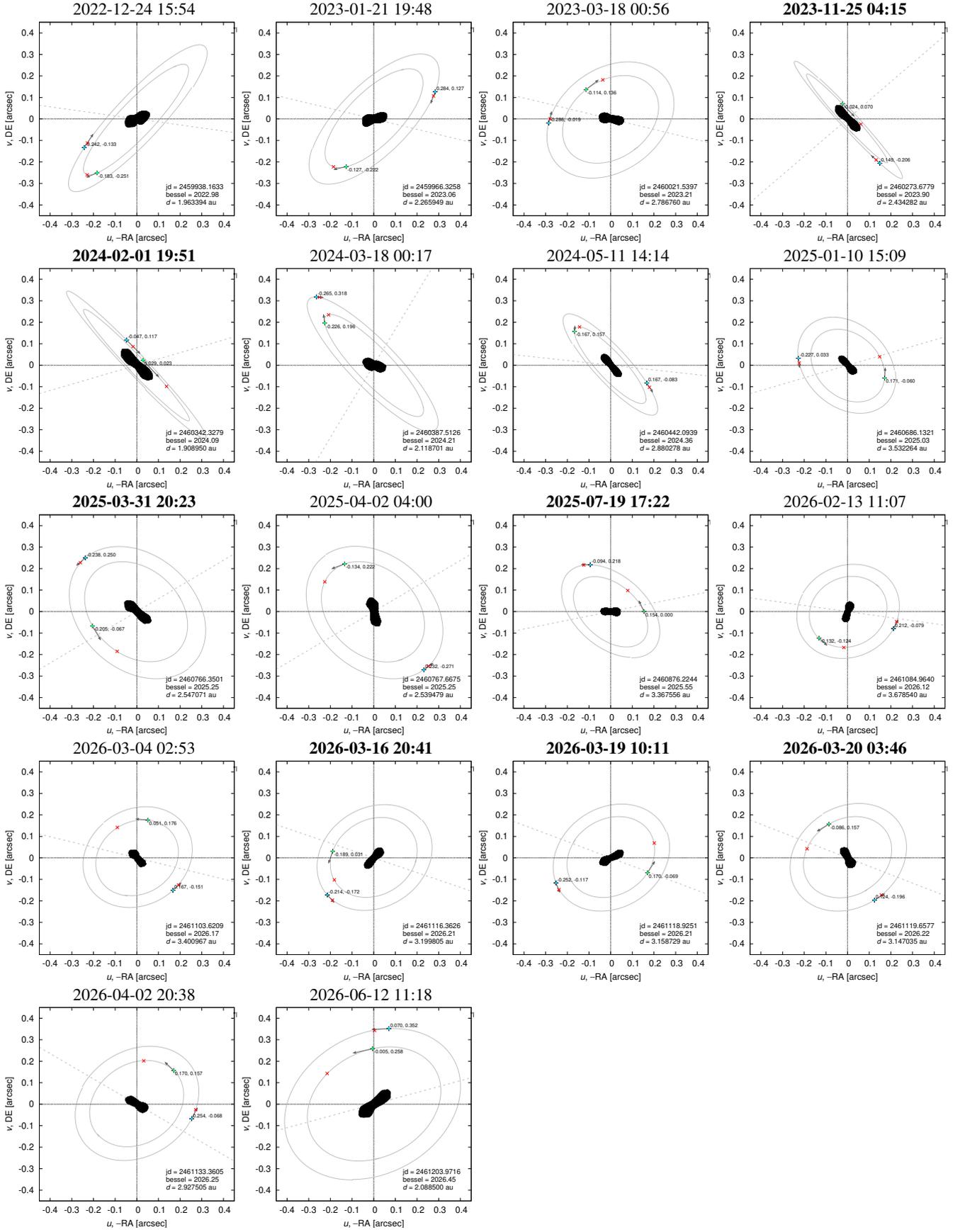

\centering
\begin{tabular}{c@{\kern1mm}c@{\kern1mm}c@{\kern1mm}c}
\kern.5cm 2022-12-24 15:54 &
\kern.5cm 2023-01-21 19:48 &
\kern.5cm 2023-03-18 00:56 &
\kern.5cm {\bf 2023-11-25 04:15} \\
\s{chi2_SKY_uv_2022.980997} &
\s{chi2_SKY_uv_2023.058103} &
\s{chi2_SKY_uv_2023.209274} &
\s{chi2_SKY_uv_2023.899606} \\
\kern.5cm {\bf 2024-02-01 19:51} &
\kern.5cm 2024-03-18 00:17 &
\kern.5cm 2024-05-11 14:14 &
\kern.5cm 2025-01-10 15:09 \\
\s{chi2_SKY_uv_2024.087563} &
\s{chi2_SKY_uv_2024.211275} &
\s{chi2_SKY_uv_2024.360713} &
\s{chi2_SKY_uv_2025.028868} \\
\kern.5cm {\bf 2025-03-31 20:23} &
\kern.5cm 2025-04-02 04:00 &
\kern.5cm {\bf 2025-07-19 17:22} &
\kern.5cm 2026-02-13 11:07 \\
\s{chi2_SKY_uv_2025.248497} &
\s{chi2_SKY_uv_2025.252104} &
\s{chi2_SKY_uv_2025.549323} &
\s{chi2_SKY_uv_2026.120833} \\
\kern.5cm 2026-03-04 02:53 &
\kern.5cm {\bf 2026-03-16 20:41} &
\kern.5cm {\bf 2026-03-19 10:11} &
\kern.5cm {\bf 2026-03-20 03:46} \\
\s{chi2_SKY_uv_2026.171914} &
\s{chi2_SKY_uv_2026.206800} &
\s{chi2_SKY_uv_2026.213816} &
\s{chi2_SKY_uv_2026.215822} \\
\kern.5cm 2026-04-02 20:38 &
\kern.5cm 2026-06-12 11:18 \\
\s{chi2_SKY_uv_2026.253339} &
\s{chi2_SKY_uv_2026.446665} &
\end{tabular}
\caption{
Predictions of Kleopatra's moons positions in the $(u,v)$ plane
for the beginning time of expected stellar occultations 2022--2026.
Our ephemerides including tides ($+$)
and without tides (\color{red}$\times$\color{black})
are plotted for comparison.
The projected orbital velocity (arrow)
and the occultation chord (dashed line) are also indicated.
If one of the alternative models is valid
(Secs.~\ref{sec:3:2}, \ref{sec:massive}),
the position of the {\em inner\/} moon will be somewhere in between.
If chords intersecting Kleopatra will be close to the inner moon,
the event is very promising (see bold dates).
}
\label{occultations_2026}
\end{figure}

\end{document}